# The calcium doping effects on the structural and electrical properties of NdFeAsO$_{0.8}$F$_{0.2}$

F. Shahbaz Tehrani, V. Daadmehr[*]

*Magnet & Superconducting Research Lab., Faculty of Physics & Chemistry, Alzahra University, Tehran 19938, Iran*

[*]Corresponding author:

Tel: (+98 21) 85692631 / (+98) 912608 9714

Fax: (+98 21) 88047861

E-mail: daadmehr@alzahra.ac.ir

URL: http:// staff.alzahra.ac.ir/daadmehr/

First author:

E-mail: tehrani66@gmail.com

**Abstract**

We have investigated experimentally how properties of NdFeAsO$_{0.8}$F$_{0.2}$ superconductor affected due to the substitution of the calcium by the neodymium. We have synthesized polycrystalline Nd$_{1-x}$Ca$_x$FeAsO$_{0.8}$F$_{0.2}$ samples with x=0, 0.01, 0.025, 0.05 and 0.1 through the one step solid-state reaction method. The X-ray diffraction patterns, refined using the MAUD software and Rietveld's method, have indicated the formation of tetragonal structure with the space group P4/nmm:2. We studied various structural parameters such as lattice parameters, bond angles, bond length, and etc. Based on the XRD data refinement, we have determined the upper limit of the calcium solubility in Nd-1111 structure and it is restricted to x≤0.05. Also, we found that the lattice parameter "a" was almost constant by increasing the calcium content, while the lattice parameter "c" and the cell volume decreased. Based on XRD data analysis, we have inference that these are due to the variations in the bond angles of As-Fe-As "α" and "β" and a decrease in the bond lengths upon increasing x. So, we have expected that the superconducting transition temperature will be sensitive to the calcium doping. Experimentally, the superconducting transition temperature was reduced from 55K (x = 0) to 48K (x = 0.01) and disappeared for other samples. On the other hand, it can be concluded from our study results about of the structural and electrical properties, which the superconducting transition temperature decreased with increasing the distortion of FeAs4-tetrahedrons from regular one. Also, based on Williamson-Hall equation, the microstrain of the samples increased upon increasing x and for Nd$_{0.95}$Ca$_{0.05}$FeAsO$_{0.8}$F$_{0.2}$ sample was grown approximately three times in comparison to the pure sample. So, it is noticeable that there is a relation between the structural properties and superconductivity in our samples.

# 1. Introduction

The discovery of iron-based superconductors (FeSCs) continues more than a decade and these superconductors have attracted the attention of researchers due to existence of their superconductivity and magnetic properties, simultaneously [1-5]. The FeSCs are one of the types of high temperature superconductors and have a layered structure like cuprate superconductors [6-8]. Among of FeSCs, the 1111-type with common formula of $ReFeAsO_{1-x}F_x$ (Re reperesnts rare-earth element atoms) have the highest superconductivity transition temperature ($T_C$) and the FeAs layers act as the superconductive planes and the Re(O/F) layers are the charge reservoirs [9-11].

Doping at the Re(O/F) layers by substitution of the electron or hole dopants changes the $T_C$ of FeSCs and the effect of these substitutions is discovered on their electrical and magnetic properties [12-14]. The effect of the calcium substitution on the structural, electrical and magnetic properties of FeSCs was discovered in the 122-compounds [15-17] but the effects of calcium doping on the 1111-type has researched slightly. A. Marcinkova et al. investigated response of the crystal structure and electronic properties to the calcium substitution in the $Nd_{1-x}Ca_xFeAsO$ compounds [18]. He found that the calcium solubility structure was restricted to x≤0.05 and the superconductivity didn't occur with an increase in the calcium content. Later, P. M. Aswathy et al. reported the effect of hole and electron dopants by co-doping substitution of $Ca^{2+}/Nd^{3+}$ and $F^-/O^{2-}$ ions, on the structural and electrical properties of NdFeAsO compound [19]. He reported the synthesis of $Nd_{1-x}Ca_xFeAsO_{1-2x}F_{2x}$ (with x=0.15, 0.2) superconductors and described that the sample with x=0.2 exhibited a maximum critical temperature ($T_C$) of 52.3K and transport critical current density ($J_C$) of 1240 A/cm$^2$ at 12 K.

Among of the 1111-type superconductors, the NdFeAsO$_{1-y}$F$_y$ compounds have the high T$_C$ (the maximum T$_C$ has been reported for y=0.2), so there are a lot of researches about the substitution effect of various ions on their structural and electrical properties [20-25]. But the calcium substitution effect on the structural and electrical properties of the NdFeAsO$_{0.8}$F$_{0.2}$ superconductor is not studied so far.

In this work for the first time, we study the effects of the Ca/Nd substitutions on polycrystalline Nd$_{1-x}$Ca$_x$FeAsO$_{0.8}$F$_{0.2}$ with x=0.0, 0.01, 0.025, 0.05, and 0.1 synthesized through one step solid-state reaction method. Specifically, we analyze the structural properties of the doped samples via X-ray diffraction (XRD) by refined with the Rietveld's method with the MAUD software. Based on the XRD data refinement, we try to determine the solubility of the calcium ions in all samples. Also, we aim to argue about the structural properties such as bond angles, bond length and microstrain comprehensively. We thus study the calcium doping effects not only regarding structural properties, but also in the sense of the superconductivity. In this paper, we experimentally try to emphasize that there is a relation between the structural properties and superconductivity in FeSCs.

## 2. Experimental

### 2.1. Preparation method

The polycrystalline Nd$_{1-x}$Ca$_x$FeAsO$_{0.8}$F$_{0.2}$ with x=0.0, 0.01, 0.025, 0.05, 0.1 were synthesized by one step solid-state reaction method, based on the preparation method that introduced by Alborzi et.al [26]. The NdFeAsO$_{0.8}$F$_{0.2}$, Nd$_{0.99}$Ca$_{0.01}$FeAsO$_{0.8}$F$_{0.2}$, Nd$_{0.975}$Ca$_{0.025}$FeAsO$_{0.8}$F$_{0.2}$, Nd$_{0.95}$Ca$_{0.05}$FeAsO$_{0.8}$F$_{0.2}$ and Nd$_{0.9}$Ca$_{0.1}$FeAsO$_{0.8}$F$_{0.2}$ samples are labeled as Nd-1111, Nd-Ca0.01, Nd-Ca0.025, Nd-Ca0.05 and Nd-Ca0.1, respectively. The neodymium powder (99.99%), arsenic pieces (99.99%), Fe$_2$O$_3$ (99.9%), FeF$_3$ (99%), Fe (99.99%) and CaF$_2$ (99%) powders were used as precursor materials. The stoichiometric amounts of these powders were mixed and grounded for 2 hours. All the processes were carried out in a glove box with Nitrogen atmosphere. The homogenous mixtures were pressed into pellets and were sealed in the evacuated quartz tubes. The sealed quartz tubes were heated up to 350ºC with 10ºC/min rate for 5 hours and followed by a same behavior at 640ºC/14 hours, 880ºC/20 hours and 1150ºC/30 hours. After this stage, the synthesized samples were again grounded, re-pelletized and sealed in the evacuated quartz tubes. Finally, the heat treatment was repeated.

### 2.2. Characterization

The X-ray diffraction patterns (XRD) of the synthesized samples were performed using a PANalytical® PW3050/60 X-ray diffractometer with Cu Kα radiation *(λ= 1.54056Å)* operated at 40kV and 40mA with a step size of 0.026°. The refinement method of Rietveld was applied with the "Material Analysis Using Diffraction" (MAUD) program (v.2.8).

The morphology of the samples was observed by a Tescan® – Mira III Field Emission Scanning Electron Microscope (FE-SEM). A four probe technique was used for electrical transport measurements and the applied DC current was 10 mA and the voltage measured with microvolt accuracy.

## 3. Results and discussion

### 3.1. Structural study

The XRD patterns of synthesized $Nd_{1-x}Ca_xFeAsO_{0.8}F_{0.2}$ (with x=0.0, 0.01, 0.025, 0.05) samples have been shown in Fig. 1. The presence of the (102), (110), (111), (112), (004), (113), (211), (114), (212), (204), (006), (221), (205), (116), (302), (310) and (215) lattice planes in the XRD patterns approves the formation of tetragonal structure with the P4/nmm:2 space group. Furthermore, the shift of the XRD-peaks to major diffraction angle with the increase in the Ca content may be attributed to this point that the sample with higher Ca content has a smaller lattice parameter (the (102) XRD-peaks are shown in Fig.2).

The XRD data of the synthesized samples have been refined by using the MAUD software with Rietveld's method for the structural analysis, occupancy number and lattice parameter calculations. Table 1 shows the before- and after-refinement values for ions positions, occupancy number and quantity of the ions in a unit cell of the synthesized samples. Based on the XRD data refinement, the formation of tetragonal structure with P4/nmm:2 space group has been confirmed in all compounds in room temperature. To investigate that how the calcium ions substitute in the synthesized Ca/Nd compounds, as well to clarify the solubility of the calcium ions, we refined the XRD data by employing the MAUD software. The agreement of the before- and after-refinement results for $Ca^{2+}$ and $Nd^{3+}$ ions in the Nd-Ca0.01, Nd-Ca0.025 and Nd-Ca0.05 compounds, illustrate the complete substitution of the calcium ions in neodymium sites.

The XRD pattern of the Nd-1111 sample shows that there are two impurity phases of FeAs and NdOF. Also, we observe the existence of $Nd_2O_3$ phase in the Nd-Ca0.01, Nd-Ca0.025 and Nd-Ca0.05 samples. These additional phases usually exist in the polycrystalline FeSCs that were reported in some previous studies [19-21, 27-30]. The peaks corresponding to $CaF_2$ and CaO are not present in the synthesized Ca/Nd samples that confirm all Ca atoms have been placed in the 1111-structure (see Fig.1). The percent volume (Vol. %) of phases for our synthesized samples are listed in Table 2 and has shown that the impurity phases have the small amounts in all samples.

The XRD pattern of the synthesized Nd-Ca0.1 sample has been displayed in Fig. 3. The phases of $FeAs_2$, FeAs, NdOF, $Nd_2O_3$ and CaAs are present in the sample.

We aim to know how many calcium ions have been entered into the 1111-structure of the synthesized Nd-Ca0.1 sample and so we need to compute the occupancy number of the calcium and the neodymium ions from XRD analysis. The MAUD analysis for the sample is shown in table 1. The maximum occupation number of the calcium ions in the Nd-Ca0.1 sample is obtained 0.049(8) and more than this amount of the calcium cannot be substituted in the neodymium sites. This issue and the presence of other peaks that can be attributed to another phase of the calcium ions (these peaks have ascribed to CaAs phase and are shown by triangles in Fig.2) imply the upper limit of the calcium solubility. So, we suggest that the solubility of the calcium ions is limited to x=0.05 in the polycrystalline $Nd_{1-x}Ca_xFeAsO_{0.8}F_{0.2}$ compounds. Although we have tried to synthesize the sample with x=0.1 content, the Nd-Ca0.05 phase (5%) forms in this sample because of the calcium solubility restriction. In a similar result, A. Marcinkova et al. [18] had specified the limit of the calcium solubility for the parent NdFeAsO sample and it was restricted to x≤0.05.

Then, we concentrate to investigate the calcium substitution effects on the Nd-1111 sample for doping values up to 0.05.

The XRD patterns of the synthesized Nd-1111, Nd-Ca0.01, Nd-Ca0.025 and Nd-Ca0.05 samples that are refined by the MAUD software are displayed in Figs. 4(a)-(d). The goodness of fit (S parameter) is characterized by $S= R_{wp}/R_{exp}$, where $R_{wp}$ is the weighted residual error and $R_{exp}$ is the expected error. These refinement parameters are listed in Table 2 and show that the refinement quality (S parameter) of our samples is good and less than 2.2.

The obtained structural parameters by employing the MAUD software for the synthesized samples are listed in Table 3. It is seen that the lattice parameter "a" is almost constant by increasing the calcium contents, but the lattice parameter "c" and the cell volume decrease. Because there isn't much difference between the ionic radiuses of $Ca^{2+}$ (1.12Å) and $Nd^{3+}$ (1.11Å), so we need to calculate other structural parameters such as bond angles, bond length, thickness of layers and the distance between layers for explaining the reason of these variations. These parameters for our samples have been listed in Tables 3 and 4. It is understood from Table 3 that the bond length of the Nd-(O/F) decreases with the increase in the calcium doping. Based on the MAUD analysis (see Table 1) and the positions of the ions that calculated after refinement, the bond length of the Nd-(O/F) determine 2.311(9) Å, 2.307(5) Å, 2.304(2) Å and 2.303(4) Å for the Nd-1111, Nd-Ca0.01, Nd-Ca0.025 and Nd-Ca0.05 samples, respectively. It is evident that the calculated Nd-(O/F) values using MAUD software are in agreement with the corresponding results (that are shown in Table 3). Also these values decrease by increasing the calcium content.

It is evident from Table 3 that the (O/F)-Nd-(O/F) angle increases with increasing the Ca content. It can be attributed to the electronegativity difference of the calcium ions in comparison to the neodymium ions. The increase in the (O/F)-Nd-(O/F) angle and the reduction in the bond length of the Nd-(O/F) lead to shrinkage of the Nd-(O/F) layer (are shown in Table 4). Also, the change of the (O/F)-Nd-(O/F) angle is effective on the variation of the As-Fe-As angles. Table 3 indicates that the bond angles of As-Fe-As "α" and "β" decreases and increases by increasing the calcium content, respectively. These issues and the decrease in the bond length of the Fe-As are leading to compression of the Fe-As layer (see Table 4). Figure 5 displays the schematic picture of the samples with bond angles of "α" and "β".

The FeAs4 tetrahedrons play an important role in the crystal structure of 1111 - structures. T. Nomura et al. [31] investigated the effect of this distortion of the FeAs4 tetrahedron from the regular value (α=109.47°) on the $T_C$. C.H. Lee et al. [32] described that the $T_C$ of FeSCs has been reduced by moving away from the regular tetrahedron value. We introduce the $\Delta_\alpha$ and $\Delta_\beta$ parameters as the relative distortion of FeAs4-tetrahedrons from regular one:

$$\Delta_\alpha = \frac{|\alpha_{regular\ tetrahedron\ angle} - \alpha_{sample}|}{\alpha_{regular\ tetrahedron\ angle}} \quad (1)$$

Where $\alpha_{regular\ tetrahedron\ angle} = \beta_{regular\ tetrahedron\ angle} = 109.47°$. These parameters are listed in Table 4 for the various calcium contents. Due to the increasing of the $\Delta_\alpha$ and $\Delta_\beta$, it is expected that the $T_C$ decreases by increasing the calcium content. By comparing the variation trend of the $T_C$, $\Delta_\alpha$ and $\Delta_\beta$ (see Table 4), T. Nomura and H. Lee's opinions are confirmed.

Furthermore, the contraction of the Fe-As and Nd-(O/F) layers lead to a reduction in the distance between these layers (that are shown in table 4) and decreasing the "c" value and the cell volume. It can be suggested that due to decreasing of the distance between layers, the energy levels are closer together. This may be leads to a change in carrier density on the superconductive planes. From the viewpoint of the BCS theory, a change in the density of states can be changed the $T_C$ [33]. As Nomura noted that the investigation of structural parameters can be a good candida to explain the superconducting effects in FeSCs. Similarly to cuprates, the displacement of the $O_4$ to the superconductive plates causes to change the carrier density and so the $T_C$ in these compounds [34, 35].

The crystallite size D and microstrain η, are calculated through Williamson-Hall equation:

$$\beta \cos\theta / \lambda = K/D + \eta \sin\theta / \lambda \quad (2)$$

Where K is Scherer's constant, β and θ are full width at half maximum (FWHM) and diffraction angle for each peak, respectively [36]. Williamson-Hall plots for $Nd_{1-x}Ca_xFeAsO_{0.8}F_{0.2}$ (x=0, 0.01, 0.025, 0.05) samples are shown in Fig. 7(a)-(d). So, the calculated microstrain and the average crystallite size of synthesized samples are given in Table 4. It can be understood from this table that increasing the calcium content leads to an enhancement of the crystallite size. Also, the microstrain of our synthesized samples increases upon increasing the calcium content and for the Nd-Ca0.05 sample is approximately grown three times in comparison to the Nd-1111 sample. According to our results obtained from structural analysis, the lattice parameter "c" and the cell volume decrease by increasing the calcium contents. So the increasing of the microstrains in our samples with the increase in the $Ca^{2+}$ ions may be due to the lattice shrinkage.

### 3.3. Morphological study

The FE-SEM images of the Nd-1111 and Nd-Ca0.01 samples have been shown in Figs. 8 (a) and (b). It can be observed from these figures that the grains are layered and grow by increasing the calcium content. Also they are randomly oriented. Also, the existence of the sharp and distinguishable layers in the synthesized samples shows high accuracy in the choosing of the important temperatures in our synthesis method.

### 3.4. Electrical characterization

The effect of the calcium doping on the electrical properties of the synthesized $Nd_{1-x}Ca_xFeAsO_{0.8}F_{0.2}$ (x=0, 0.01, 0.025 and 0.05) samples is studied through resistivity measurements by 4-prob technique. The temperature dependence of $R/R_0$ for our synthesized samples has been displayed in Fig. 9(a)-(b). For the Nd-1111 sample, the electrical resistivity gradually declines by decreasing temperature and then the superconductivity transition occurs at $T_C^{mid} = 55K$. The Nd-Ca0.01 sample represents the structural transition at 125K and then the electrical resistivity slowly decreases by cooling. The superconductivity transition of this sample happens at $T_C^{mid} = 48K$. The existence of structural transition suggests that the superconductivity appears in the orthorhombic phase [32]. By more increasing the calcium content, the superconductivity suppresses in other samples: I) the structural transition occurs at 145K for the Ca0.025 sample and the electrical resistivity linearly decreases by cooling and it behaves as a metallic compound. II) The Nd-Ca0.05 sample is not a superconductor and it displays a semiconducting behavior. Therefore, based on our electrical measurements it can be said that:

1) With an increase in the calcium content, the structural transition temperature shifts to higher temperature. Such behavior are mentioned by S. Matsuishi et al. [37] for the $CaTM_{1-x}Fe_xAsF$ (TM=Ni, Co, Mn) compounds.

2) It seems that the $T_C$ decreases by substitution of $Ca^{2+}/Nd^{3+}$ ions. By comparing the results obtained from our structural and electrical measurements, it can be said that the $T_C$ decreases with increasing the relative distortion of FeAs4-tetrahedrons (see Table 4). So the angular changes in the superconductivity planes influence on the superconductivity transition. T. Nomura et al. [31] studied the effect of this distortion on the $T_C$ in $CaFe_{1-x}Co_xAsF$ samples and C.H. Lee et al. [32] concluded that the $T_C$ of $LnFeAsO_{1-y}$ compounds have been reduced by moving away from the regular tetrahedron value. Our results are in good agreement with these works. It is Similar to the cuprates [38].

In addition, Table 4 indicates that the increasing of the microstrain causes to suppression of the superconductivity in our samples upon increasing the calcium content. So, it can be suggested that the study of the microstrain introduce as a global factor for the variations of the $T_C$.

## 4. Conclusion

We have synthesized polycrystalline $Nd_{1-x}Ca_xFeAsO_{0.8}F_{0.2}$ samples with x=0, 0.01, 0.025, 0.05 and 1 through the one step solid-state reaction method. We have investigated experimentally structural and electrical properties. The X-ray diffraction patterns were refined using the MAUD software and Rietveld's method.

Based on the XRD data refinement, all the synthesized samples had the tetragonal structures with the space group P4/nmm:2. By synthesizing the several samples, based on the MAUD analysis, the calcium solubility limit was obtained 0.05. The decrement of the lattice parameter "c" was explained through the decrease and increase of the bond angles "α" and "β", respectively and as a result the reduction in the thickness of the layers. Due to the increasing of the relative distortions "$\Delta_\alpha$" and "$\Delta_\beta$" the $T_C$ decreased by increasing the calcium content. Our structural calculations showed that the microstrain of the $Nd_{1-x}Ca_xFeAsO_{0.8}F_{0.2}$ samples was increased by substitution of the calcium content which can be attributed to the lattice constriction. This result was consistent with the reduction of the lattice parameter. The superconducting transition temperature is reduced from 55K (x = 0) to 48K (x = 0.01) and disappeared for other samples. The increasing of the microstrain, "$\Delta_\alpha$" and "$\Delta_\beta$" caused to suppression of the superconductivity in our samples with the increase in the calcium content. So, we emphasize that there is a relation between the structural properties and superconductivity in the $Nd_{1-x}Ca_xFeAsO_{0.8}F_{0.2}$ samples, as the similar results were obtained by T. Nomura in $CaFe_{1-x}Co_xAsF$ and C.H. Lee in $LnFeAsO_{1-y}$ compounds.

## Acknowledgements

The authors acknowledge the Alzahra University.

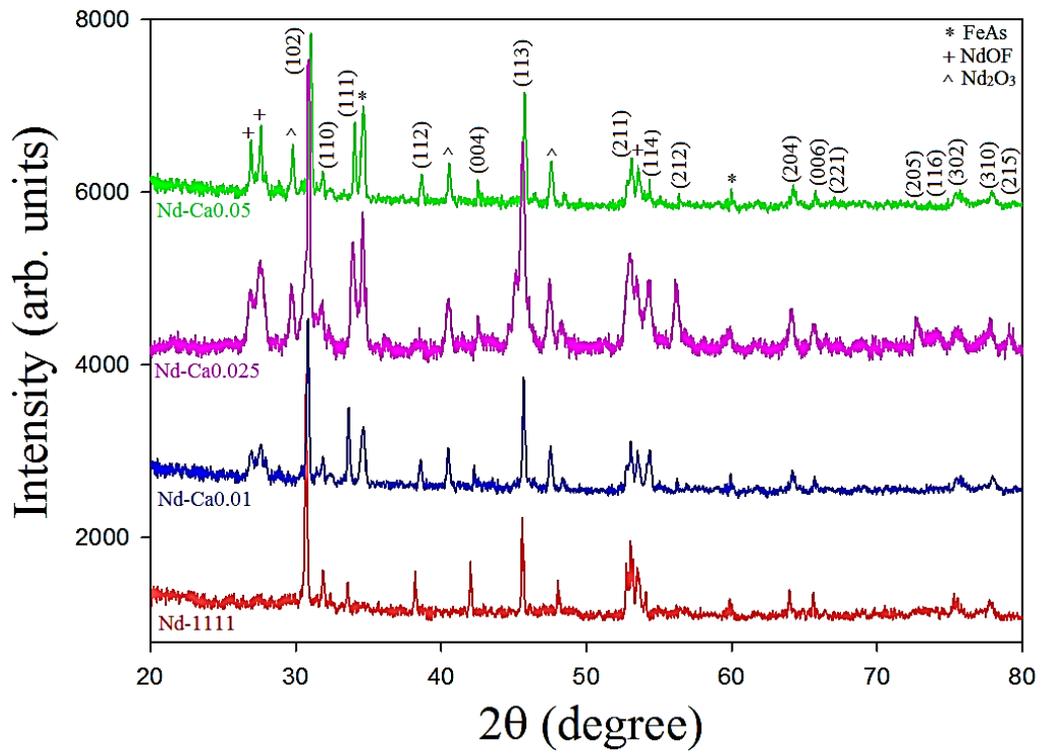

**Fig. 1. XRD patterns of synthesized Nd$_{1-x}$Ca$_x$FeAsO$_{0.8}$F$_{0.2}$ samples**

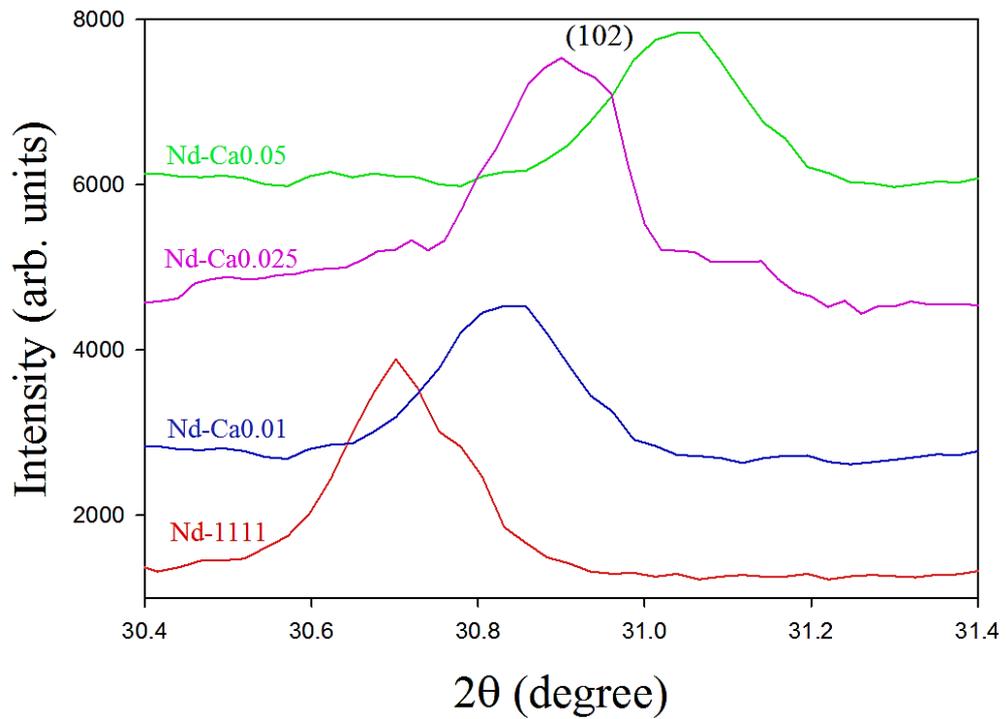

**Fig.2. Enlarged view of the (102) peaks showing the shift to the larger angle with higher Ca content**

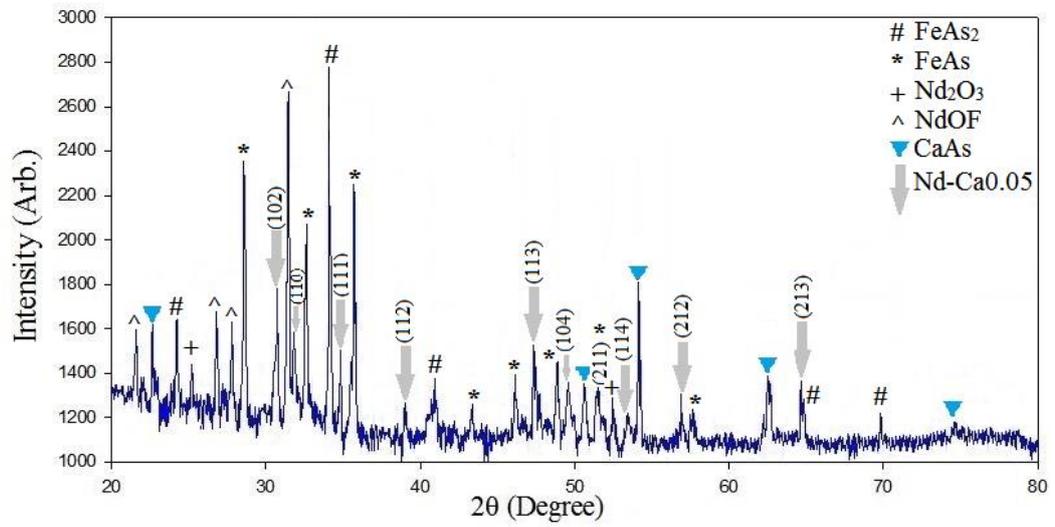

**Fig. 3. XRD patterns of the synthesized Nd-Ca0.1 sample**

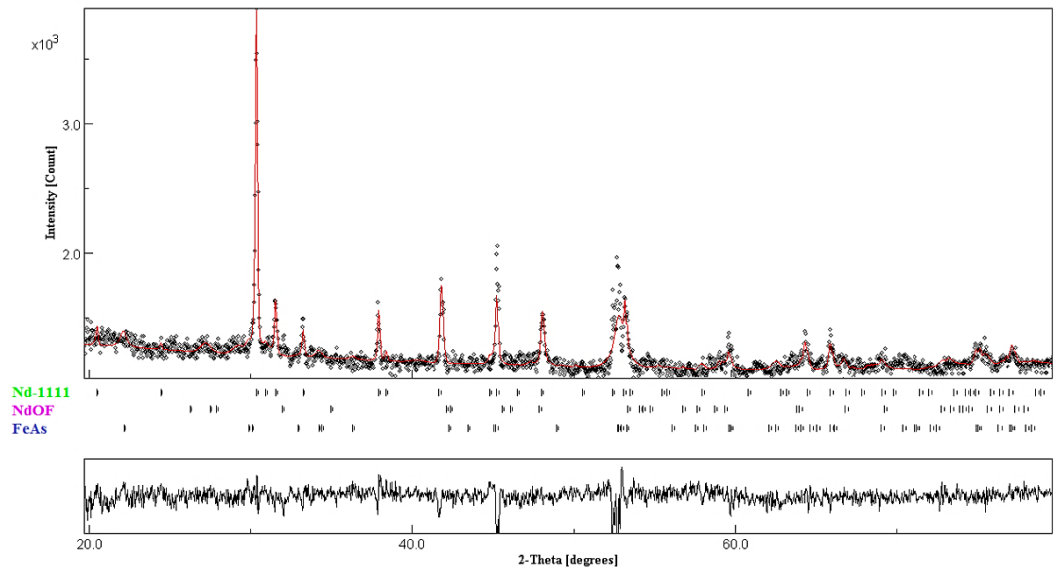

**Fig. 4(a). XRD pattern refinements using MAUD software for the Nd-1111 sample ( •: experimental data, upper solid line: calculated pattern, lower solid line: subtracted pattern)**

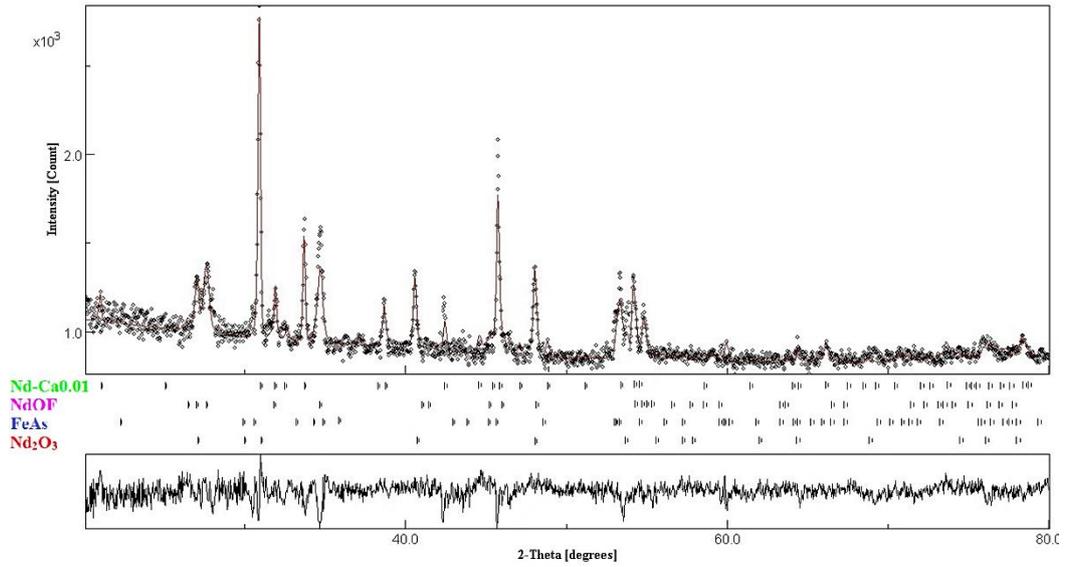

**Fig. 4(b). XRD pattern refinements using MAUD software for the Nd-Ca0.01 sample ( •: experimental data, upper solid line: calculated pattern, lower solid line: subtracted pattern)**

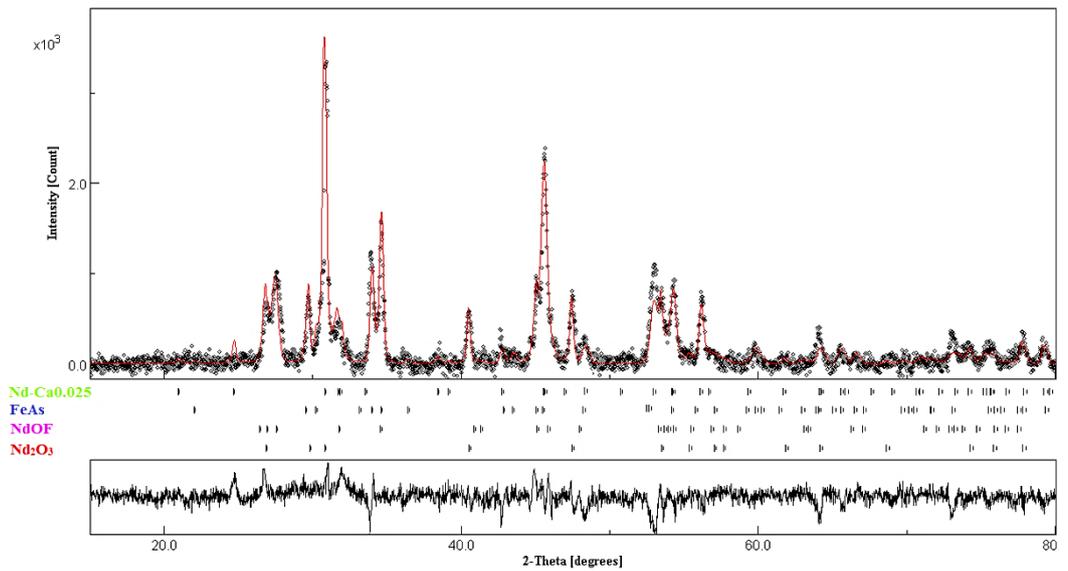

**Fig. 4(c). XRD pattern refinements using MAUD software for the Nd-Ca0.05 sample ( •: experimental data, upper solid line: calculated pattern, lower solid line: subtracted pattern)**

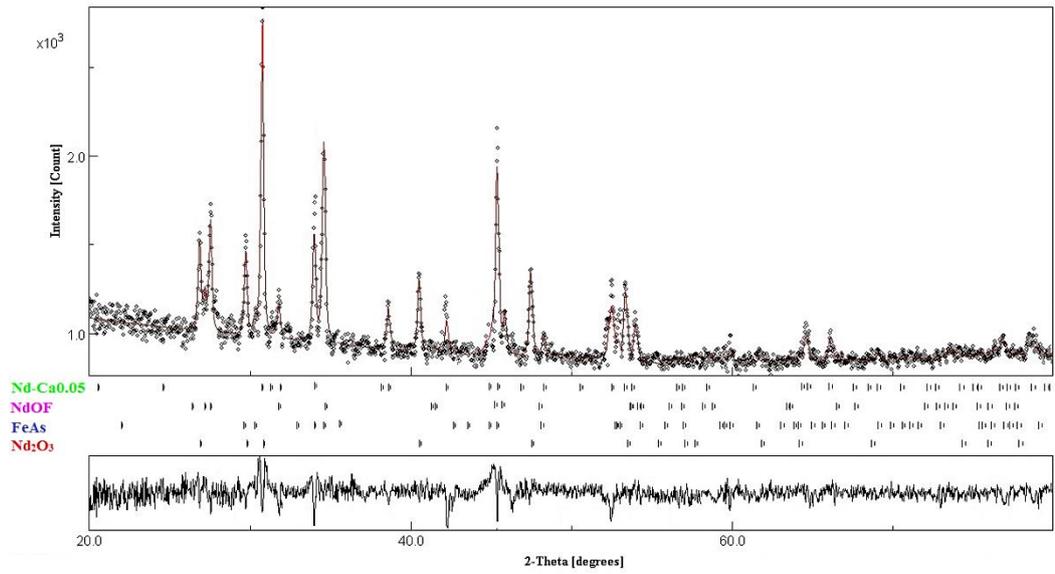

**Fig. 4(d). XRD pattern refinements using MAUD software for the Nd-Ca0.05 sample ( ●: experimental data, upper solid line: calculated pattern, lower solid line: subtracted pattern)**

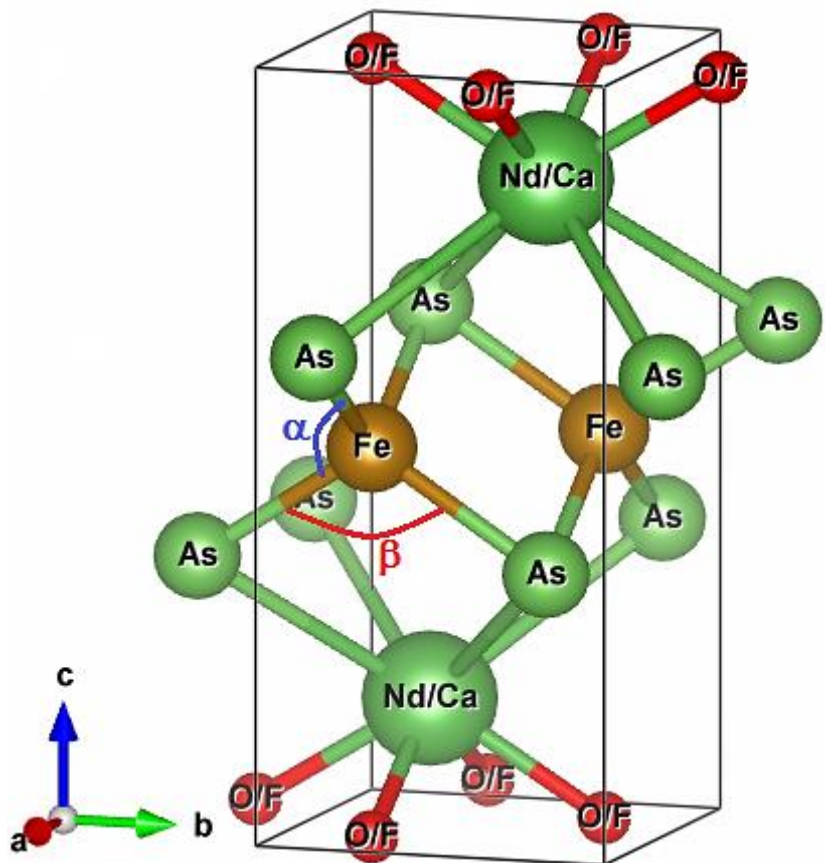

**Fig.5. Schematic picture of $Ni_{1-x}Ca_xFeAsO_{0.8}F_{0.2}$ samples**

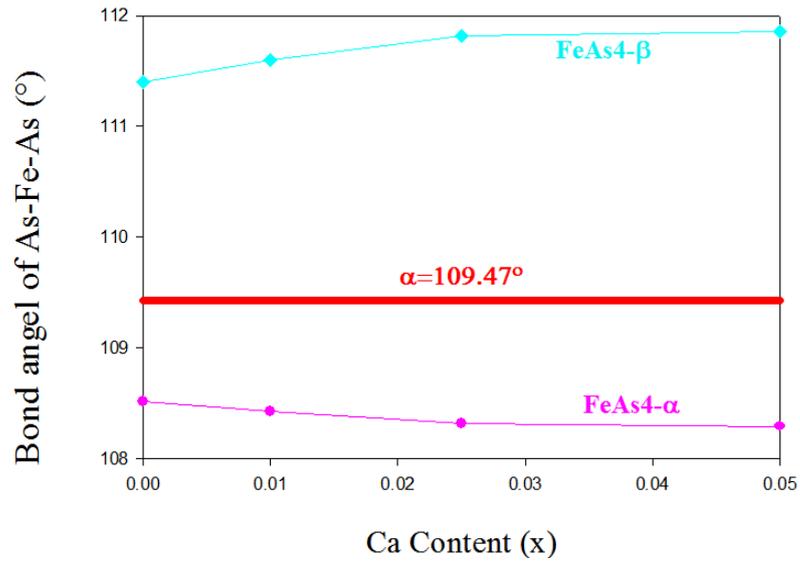

**Fig.6. Variation of bond angles of FeAs4-tetrahedron with respect to the Ca content in the synthesized $Nd_{1-x}Ca_xFeAsO_{0.8}F_{0.2}$ (x-0, 0.01, 0.025, 0.05) samples (red line indicates the bond angle of regular FeAs4-tetrahedron)**

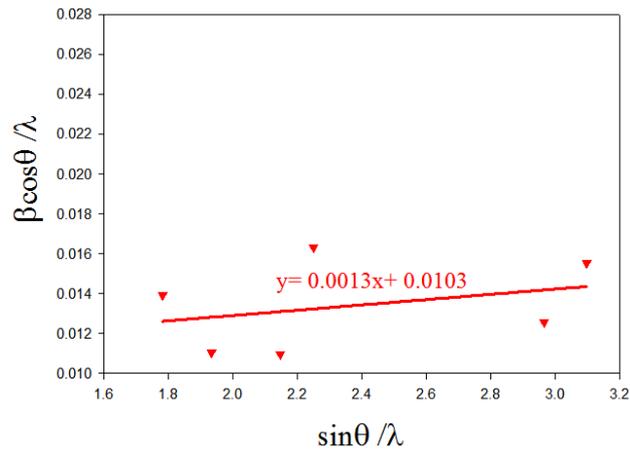

**Fig. 7(a). Williamson-Hall plot of the synthesized Nd-1111**

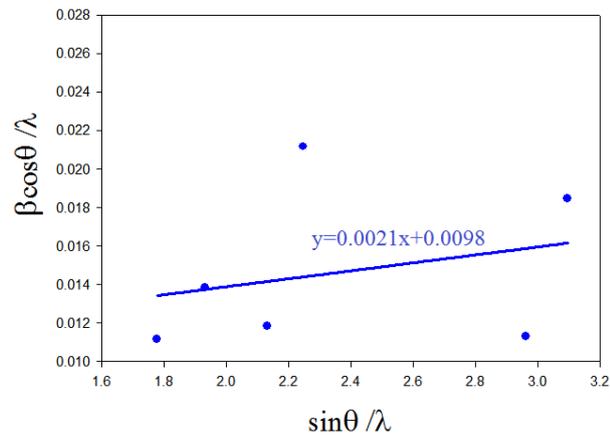

**Fig. 7(b). Williamson-Hall plot of the synthesized Nd-Ca0.01**

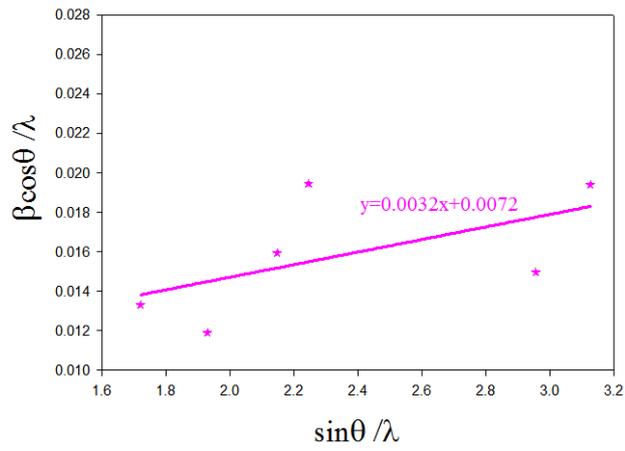

**Fig. 7(c). Williamson-Hall plot of the synthesized Nd-Ca0.025**

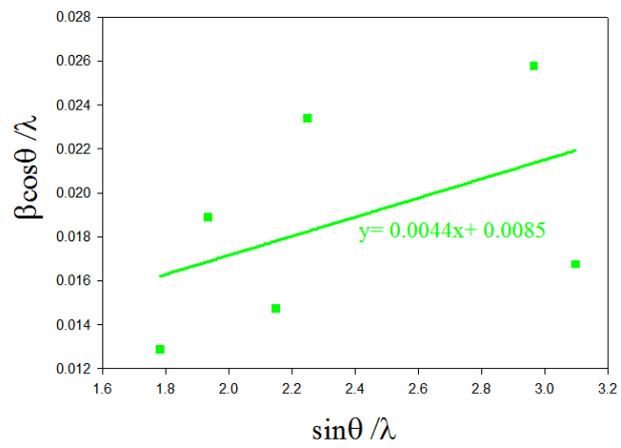

**Fig. 7(d). Williamson-Hall plot of the synthesized Nd-Ca0.05**

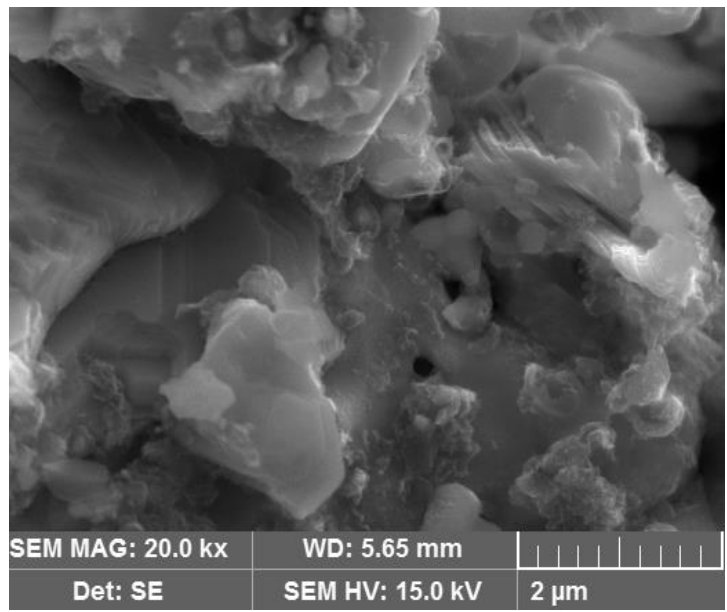

**Fig. 8(a). FE-SEM image of the synthesized Nd-1111**

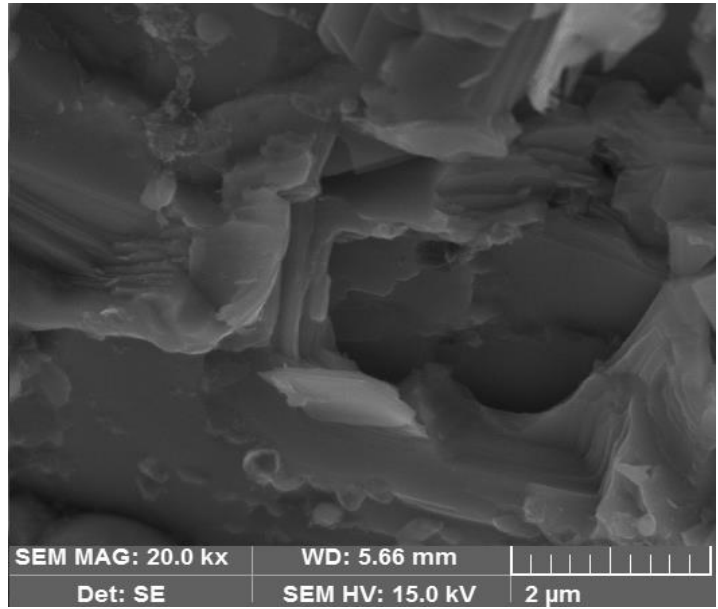

**Fig. 8(b). FE-SEM image of the synthesized Nd-Ca0.01**

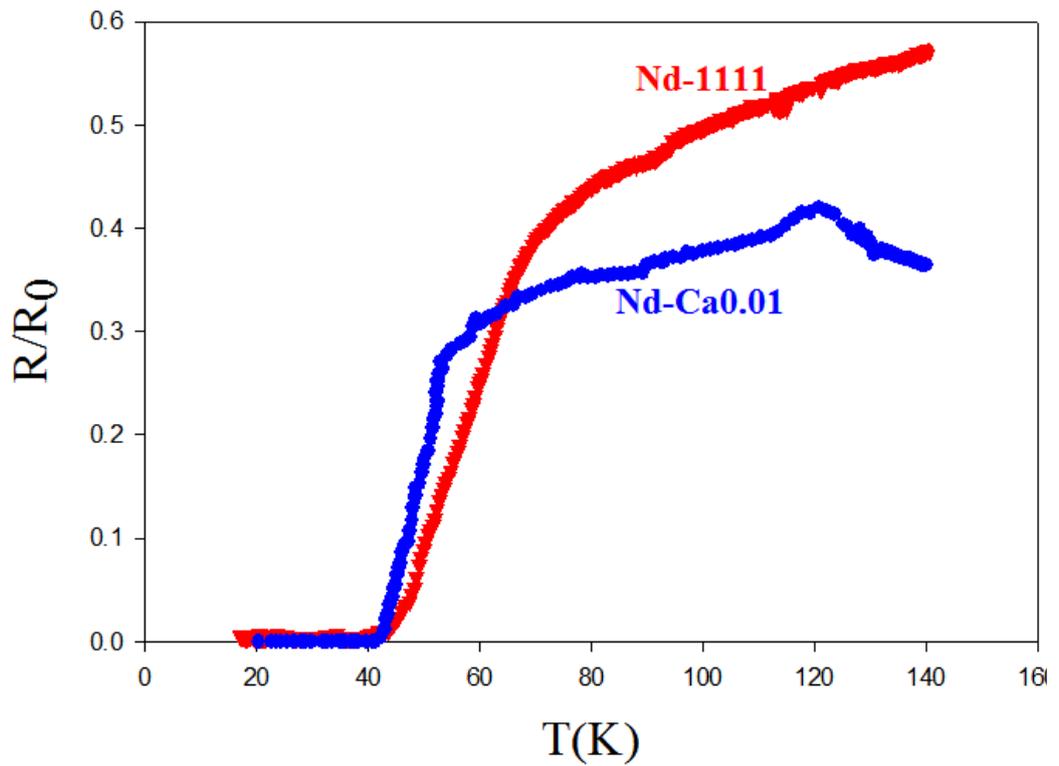

**Fig. 9(a). Temperature dependence of $R/R_0$ for the Nd-1111 and Nd-Ca0.01 samples**

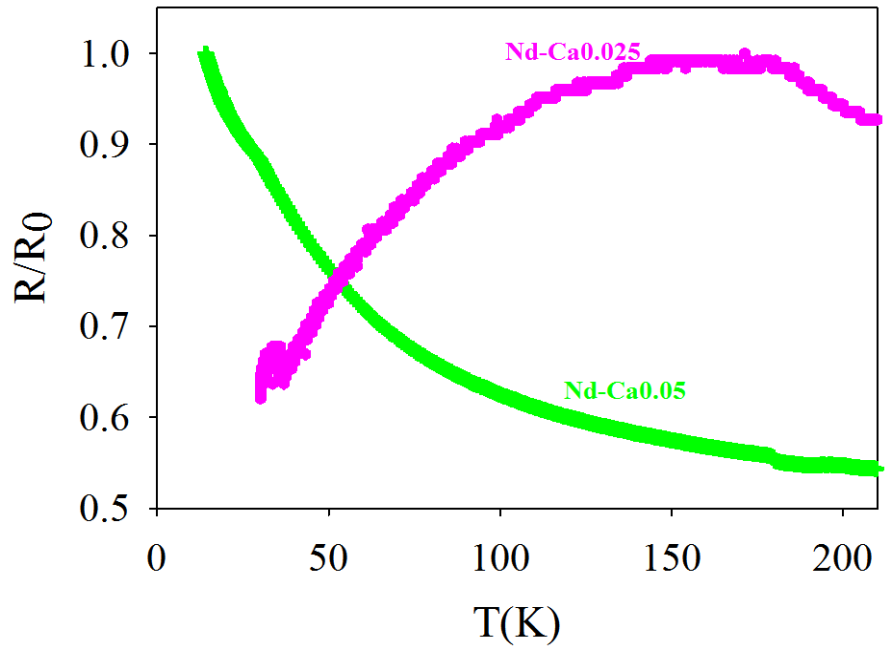

**Fig. 9(b). Temperature dependence of R/R$_0$ for the Nd-Ca0.025 and Nd-Ca0.05 samples**

| Sample name | Structural parameters before refinement | | | | | Structural parameters after refinement | | | | |
|---|---|---|---|---|---|---|---|---|---|---|
| | Ions | Position | | | Occupancy | Ions | Position | | | Occupancy |
| | | x | y | z | | | x | y | z | |
| **Nd-1111** | $Nd^{3+}$ | 0.25 | 0.25 | 0.1385 | 1 | $Nd^{3+}$ | 0.2499 | 0.2499 | 0.1381 | 0.999(8) |
| | $Fe^{2+}$ | 0.75 | 0.25 | 0.5 | 1 | $Fe^{2+}$ | 0.75 | 0.25 | 0.5 | 1 |
| | As | 0.25 | 0.25 | 0.6574 | 1 | As | 0.25 | 0.25 | 0.6574 | 1 |
| | $O^{2-}$ | 0.75 | 0.25 | 0 | 0.8 | $O^{2-}$ | 0.75 | 0.25 | 0 | 0.8 |
| | $F^{-}$ | 0.75 | 0.25 | 0 | 0.2 | $F^{-}$ | 0.75 | 0.25 | 0 | 0.2 |
| **Nd-Ca0.01** | $Ca^{2+}$ | 0.25 | 0.25 | 0.1385 | 0.01 | $Ca^{2+}$ | 0.2498 | 0.2499 | 0.1379 | 0.010(5) |
| | $Nd^{3+}$ | 0.25 | 0.25 | 0.1385 | 0.99 | $Nd^{3+}$ | 0.2497 | 0.2499 | 0.1374 | 0.986(6) |
| | $Fe^{2+}$ | 0.75 | 0.25 | 0.5 | 1 | $Fe^{2+}$ | 0.75 | 0.25 | 0.5 | 1 |
| | As | 0.25 | 0.25 | 0.6574 | 1 | As | 0.25 | 0.25 | 0.6574 | 1 |
| | $O^{2-}$ | 0.75 | 0.25 | 0 | 0.8 | $O^{2-}$ | 0.75 | 0.25 | 0 | 0.8 |
| | $F^{-}$ | 0.75 | 0.25 | 0 | 0.2 | $F^{-}$ | 0.75 | 0.25 | 0 | 0.2 |
| **Nd-Ca0.025** | $Ca^{2+}$ | 0.25 | 0.25 | 0.1385 | 0.025 | $Ca^{2+}$ | 0.2498 | 0.2498 | 0.1377 | 0.023(9) |
| | $Nd^{3+}$ | 0.25 | 0.25 | 0.1385 | 0.975 | $Nd^{3+}$ | 0.2497 | 0.2496 | 0.1372 | 0.974(6) |
| | $Fe^{2+}$ | 0.75 | 0.25 | 0.5 | 1 | $Fe^{2+}$ | 0.75 | 0.25 | 0.5 | 1 |
| | As | 0.25 | 0.25 | 0.6574 | 1 | As | 0.25 | 0.25 | 0.6574 | 1 |
| | $O^{2-}$ | 0.75 | 0.25 | 0 | 0.8 | $O^{2-}$ | 0.75 | 0.25 | 0 | 0.8 |
| | $F^{-}$ | 0.75 | 0.25 | 0 | 0.2 | $F^{-}$ | 0.75 | 0.25 | 0 | 0.2 |
| **Nd-Ca0.05** | $Ca^{2+}$ | 0.25 | 0.25 | 0.1385 | 0.05 | $Ca^{2+}$ | 0.2497 | 0.2499 | 0.1376 | 0.048(9) |
| | $Nd^{3+}$ | 0.25 | 0.25 | 0.1385 | 0.95 | $Nd^{3+}$ | 0.2496 | 0.2496 | 0.1370 | 0.950(6) |
| | $Fe^{2+}$ | 0.75 | 0.25 | 0.5 | 1 | $Fe^{2+}$ | 0.75 | 0.25 | 0.5 | 1 |
| | As | 0.25 | 0.25 | 0.6574 | 1 | As | 0.25 | 0.25 | 0.6574 | 1 |
| | $O^{2-}$ | 0.75 | 0.25 | 0 | 0.8 | $O^{2-}$ | 0.75 | 0.25 | 0 | 0.8 |
| | $F^{-}$ | 0.75 | 0.25 | 0 | 0.2 | $F^{-}$ | 0.75 | 0.25 | 0 | 0.2 |
| **Nd-Ca0.1*** | $Ca^{2+}$ | 0.25 | 0.25 | 0.1385 | 0.05 | $Ca^{2+}$ | 0.2496 | 0.2499 | 0.1379 | 0.049(8) |
| | $Nd^{3+}$ | 0.25 | 0.25 | 0.1385 | 0.95 | $Nd^{3+}$ | 0.2495 | 0.2496 | 0.1371 | 0.949(7) |
| | $Fe^{2+}$ | 0.75 | 0.25 | 0.5 | 1 | $Fe^{2+}$ | 0.75 | 0.25 | 0.5 | 1 |
| | As | 0.25 | 0.25 | 0.6574 | 1 | As | 0.25 | 0.25 | 0.6574 | 1 |
| | $O^{2-}$ | 0.75 | 0.25 | 0 | 0.8 | $O^{2-}$ | 0.75 | 0.25 | 0 | 0.8 |
| | $F^{-}$ | 0.75 | 0.25 | 0 | 0.2 | $F^{-}$ | 0.75 | 0.25 | 0 | 0.2 |

Table 1. Structural parameters before and after refinement from MAUD analysis for the synthesized samples (∗ although we have tried to synthesize the sample with the stoichiometric amount x=0.1, the Nd-Ca0.05 phase forms in this sample because of the calcium solubility restriction. In addition, the percent volume of the Nd-Ca0.05 phase is 5 %.)

| Sample name | $R_{wp}$ (%) | $R_p$ (%) | $R_{exp}$ (%) | S |
|---|---|---|---|---|
| Nd-1111 | 5.71(3) | 4.55(2) | 3.32(3) | 1.71(9) |
| Nd-Ca0.01 | 6.00(1) | 4.64(3) | 3.21(4) | 1.86(7) |
| Nd-Ca0.025 | 7.71(5) | 4.28(1) | 3.53(1) | 2.18(5) |
| Nd-Ca0.05 | 4.65(6) | 4.51(9) | 3.41(1) | 1.36(5) |
| Nd-Ca0.1 | 8.02(2) | 4.98(9) | 3.82(1) | 2.09(9) |

Table 2. The parameters for the calculation of pattern fitness for the synthesized samples

| Ca content (x) | Vol. %. of phases | | | | Lattice parameters | | Cell volume (Å³) | Bond length (Å) | | Bond angle (°) | | |
|---|---|---|---|---|---|---|---|---|---|---|---|---|
| | pure | FeAs | NdOF | Nd$_2$O$_3$ | a (Å) | c (Å) | | Fe-As | Nd-(O/F) | As-Fe-As | | (O/F)-Nd-(O/F) |
| | | | | | | | | | | α | β | |
| x=0 | 81.5 | 10.7 | 7.8 | 0 | 3.96(7) | 8.59(6) | 135.27(6) | 2.401(0) | 2.313(4) | 108.51(5) | 111.40(2) | 116.95(0) |
| x=0.01 | 77.0 | 9.8 | 8.7 | 4.5 | 3.96(4) | 8.55(8) | 134.47(4) | 2.396(4) | 2.309(4) | 108.42(4) | 111.60(7) | 117.13(8) |
| x=0.025 | 74.2 | 11.4 | 8.5 | 5.9 | 3.96(3) | 8.52(1) | 133.82(5) | 2.392(7) | 2.306(3) | 108.31(4) | 111.81(7) | 117.34(5) |
| x=0.05 | 73.8 | 11.7 | 7.7 | 6.7 | 3.96(1) | 8.51(0) | 133.51(7) | 2.390(9) | 2.304(7) | 108.29(4) | 111.86(7) | 117.38(6) |

Table 3. Different structural parameters for the various calcium contents from MAUD analysis

| Ca content (x) | Thickness of Fe-As layer (Å) | Thickness of Nd-(O/F) layer (Å) | Distance between Fe-As and Nd-(O/F) layers | Williamson-Hall equation | Crystalline size (nm) | Microstrain η (%) | Relative distortion Δ (%) | | T$_C$ (K) | T$_S$ (K) |
|---|---|---|---|---|---|---|---|---|---|---|
| | | | | | | | Δ$_α$ | Δ$_β$ | | |
| x=0 | 1.353(2) | 1.209(6) | 1.735(8) | y= 0.0013x+ 0.0103 | 97.09±0.04 | 0.13±0.02 | 0.87(6) | 1.76(3) | 55 | - |
| x=0.01 | 1.347(1) | 1.204(3) | 1.727(6) | y=0.0021x+0.0098 | 102.41±0.06 | 0.21±0.04 | 0.95(9) | 1.94(5) | 48 | 125 |
| x=0.025 | 1.341(1) | 1.199(1) | 1.720(3) | y=0.0032x+0.0072 | 120.48±0.05 | 0.32±0.02 | 1.05(6) | 2.14(4) | - | 145 |
| x=0.05 | 1.339(4) | 1.197(5) | 1.718(1) | y= 0.0044x+ 0.0085 | 117.65±0.09 | 0.44±0.09 | 1.07(7) | 2.18(3) | - | - |

Table 4. Some structural and superconducting parameters for the various calcium contents